\documentclass[apj]{emulateapj}
\usepackage{amsmath}

\usepackage{epsf}

\shortauthors{Belokurov et al.}
\shorttitle{Field of Streams}

\eqsecnum

\renewcommand\email\texttt

\def\spose#1{\hbox to 0pt{#1\hss}}
\def\lta{\mathrel{\spose{\lower 3pt\hbox{$\sim$}}
    \raise 2.0pt\hbox{$<$}}}
\def\gta{\mathrel{\spose{\lower 3pt\hbox{$\sim$}}
    \raise 2.0pt\hbox{$>$}}}

\begin{document} 

\slugcomment{\sc to be submitted to \it the Astrophysical Journal}
\shorttitle{\sc The Field of Streams}
\shortauthors{\sc V.~Belokurov et al.}

\title{The Field of Streams: Sagittarius and its Siblings}
\author{V. Belokurov\altaffilmark{1},
D.\ B. Zucker\altaffilmark{1}, 
N.\ W. Evans\altaffilmark{1}, 
G. Gilmore\altaffilmark{1}, 
S. Vidrih\altaffilmark{1}, 
D.\ M. Bramich\altaffilmark{1}, 
H.\ J.\ Newberg\altaffilmark{2},
R.\ F.\ G.\ Wyse\altaffilmark{3},
M.\ J. Irwin\altaffilmark{1},  
M. Fellhauer\altaffilmark{1},
P.\ C. Hewett\altaffilmark{1}, 
N.\ A. Walton\altaffilmark{1}, 
M.\ I. Wilkinson\altaffilmark{1},
N. Cole\altaffilmark{2},
B. Yanny\altaffilmark{4},
C.\ M.\ Rockosi\altaffilmark{5}
T.\ C. Beers\altaffilmark{6},
E.\ F.\ Bell\altaffilmark{7},
J. Brinkmann\altaffilmark{8},
\v{Z}.\ Ivezi\'{c}\altaffilmark{9},
R. Lupton\altaffilmark{10}}

\altaffiltext{1}{Institute of Astronomy, University of Cambridge,
Madingley Road, Cambridge CB3 0HA, UK;\email{vasily,zucker,nwe@ast.cam.ac.uk}}
\altaffiltext{2}{Rensselaer Polytechnical Institute, Troy, NY 12180}
\altaffiltext{3}{The Johns Hopkins University, 3701 San Martin Drive,
Baltimore, MD 21218}
\altaffiltext{4}{Fermi National Accelerator Laboratory, P.O. Box 500,
Batavia, IL 60510}
\altaffiltext{5}{Lick Observatory, University of California, Santa Cruz, CA 95064.}
\altaffiltext{6}{Department of Physics and Astronomy, CSCE: Center for
the Study of Cosmic Evolution, and JINA: Joint Institute for Nuclear
Astrophysics, Michigan State University, East Lansing, MI 48824}
\altaffiltext{7}{Max Planck Institute for Astronomy, K\"{o}nigstuhl
17, 69117 Heidelberg, Germany}
\altaffiltext{8}{Apache Point Observatory, P.O. Box 59, Sunspot, NM
88349}
\altaffiltext{9}{Astronomy Department, University of Washington, Box 351580,
Seattle WA 98195-1580}
\altaffiltext{10}{Princeton University Observatory, Princeton, NJ
08544}

\begin{abstract}
We use Sloan Digital Sky Survey (SDSS) Data Release 5 (DR5)
$u,g,r,i,z$ photometry to study Milky Way halo substructure in the
area around the North Galactic Cap. A simple color cut ($g-r < 0.4$)
reveals the tidal stream of the Sagittarius dwarf spheroidal, as well
as a number of other stellar structures in the field. Two branches (A
and B) of the Sagittarius stream are clearly visible in an
RGB-composite image created from 3 magnitude slices, and there is also
evidence for a still more distant wrap behind the A branch. A
comparison of these data with numerical models suggests that the shape
of the Galactic dark halo is close to spherical.
\end{abstract}


\section{Introduction}

Stellar streams in the Milky Way halo produced by the accretion of
smaller galaxies are a standard prediction of hierarchical merging
cosmogonies~\citep[e.g., ][ and references therein]{LB95}.  The most
spectacular example is the disrupting Sagittarius dwarf spheroidal
(Sgr dSph), originally discovered by Ibata, Gilmore \& Irwin
(1995). It has a heliocentric distance of $\sim 25$ kpc and is
centered at Galactic coordinates of $\ell = 5.6^\circ$ and $b =
-14.0^\circ$ \citep{Ib97}.  It is dominated by an intermediate age
population (between 6 and 9 Gyrs, Bellazzini et al. 2006), but there
is evidence for a much older population ($>$ 10 Gyrs) as
well~\citep{Mo03}. The metallicity [Fe/H] ranges from very metal-poor
(as low as -2 based on the globular clusters) up to approximately
solar, with probably a mean of $\sim -0.5$ \citep[][ and references
therein]{Mo05}. It was realised early on that there was some tidal
debris in the neighbourhood of the Sgr dSph \citep{Ib97,Ma99} and that
the distribution of this material traced the Sgr dSph
orbit. Subsequently, \citet{Ya00} used Sloan Digital Sky Survey (SDSS)
first-year commissioning data to identify an over-density of blue
A-type stars in two strips located at ($\ell,b,R$) = ($350^\circ$,
$50^\circ$, 46 kpc) and ($157^\circ$, $-58^\circ$, 33 kpc), which were
then matched with the Sgr stream~\citep{Ib01}. Likewise, \citet{Iv00}
noticed that clumps of RR Lyrae stars in SDSS commissioning data lay
along the Sgr stream's orbit.

The best panorama of the Sgr stream to date was obtained by
\cite{Ma03} using M giants selected from the Two Micron All-Sky Survey
(2MASS). They saw the trailing tidal tail very clearly in the Southern
Galactic hemisphere, as well as part of the leading arm reaching
towards the North Galactic Cap. Here, we use SDSS Data Release 5 (DR5)
to provide a picture of the leading arm of the Sgr stream in the
vicinity of the North Galactic Cap with remarkable clarity, together
with a number of other notable stellar structures in the field.
%with possible multiple wraps in the vicinity of the North Galactic
%Cap.

%
\begin{figure*}
\begin{center}
%\vskip8cm
\includegraphics[height=7.0cm]{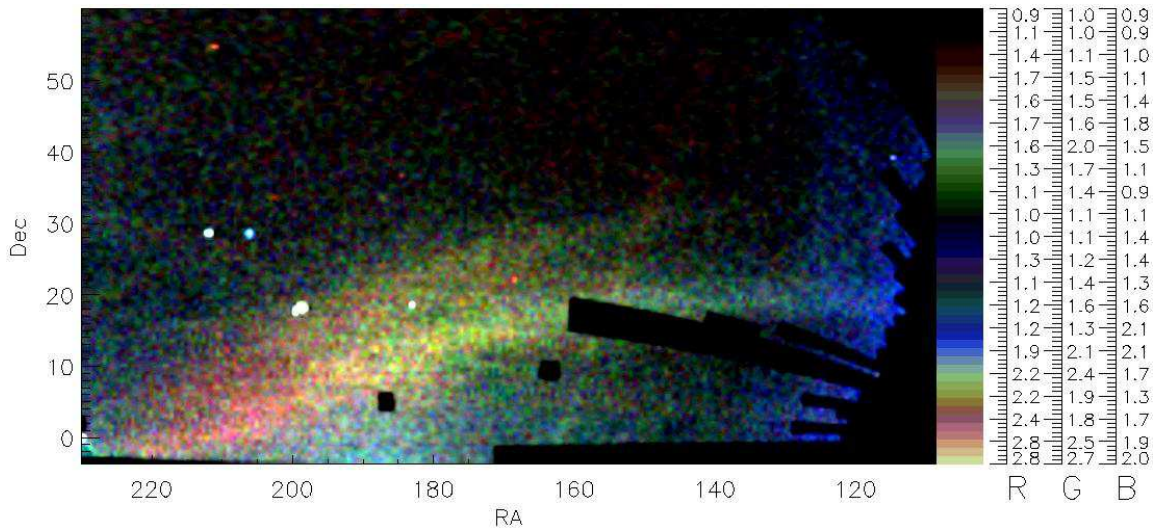}
%\plotone{f1jpg.ps}
\caption{The spatial density of SDSS stars with $g-r < 0.4$ around the
North Galactic Cap in equatorial coordinates, binned $0.5 \times 0.5$
arcdegrees. The color plot is an RGB composite with blue for the most
nearby stars with $20.0 < r \le 20.66$, green for stars with $20.66 <
r \le 21.33$ and red for the most distant stars with $21.33 < r \le
22.0$.  Note the bifurcation in the stream starting at $\alpha \approx
180^\circ$. Further structure that is visible includes the Monoceros
Ring at $\alpha \approx 120^\circ$, and a new thin stream at
$150^\circ \lta \alpha \lta 160^\circ$ and $ 0^\circ \lta \delta \lta
30^\circ$.  The color bar shows a palette of 50 representative colors
labeled according to the stellar density (in units of 100 stars per
square degree) in each of red, green and blue components. The
displayed density ranges are 102 to 330 (red), 107 to 304 (green) and
98 to 267 (blue).}
\label{fig:sag}
\end{center}
%\vskip-0.5cm
\end{figure*}
\begin{figure*}
\begin{center}
%\vskip5cm
\includegraphics[height=5.5cm]{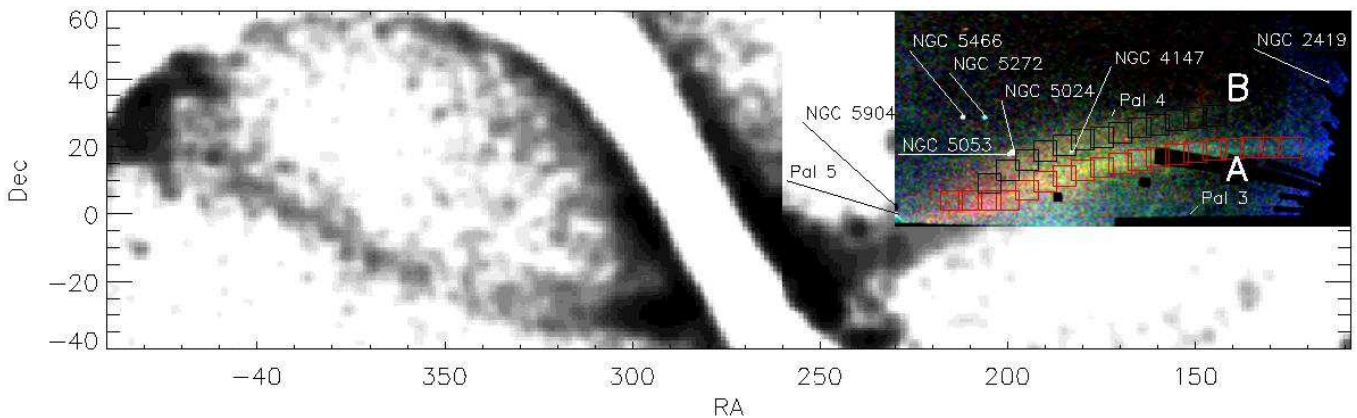}
%\plotone{f2jpg.ps}
\caption{A panoramic view of the Sgr stream, obtained by combining the
2MASS M giants of Majewski et al. (2003) with the SDSS
stars. Marked on the figure are branches A and B of the stream,
together with some of the (possibly associated) globular
clusters. Shown in red and black are the on-stream fields used in the
analysis of Section 3 (see main text).}
\label{fig:sagnew}
\end{center}
%\vskip-0.5cm
\end{figure*}
\begin{table*}
\tiny
\begin{center}
\caption{Locations of the on-stream fields along branch A ($\alpha,
\delta_A$) and branch B ($\alpha, \delta_B$). The companion off-stream
fields for branch A have the same right ascension but are offset in
declination by $+20^\circ$ for $\alpha > 150^\circ$ and by $-12^\circ$
otherwise. For branch B, the off-stream fields are offset by
$+13^\circ$ in declination.}
\begin{tabular}{|l|c|c|c|c|c|c|c|c|c|c|c|c|c|c|c|c|c|c|c|}\hline
Field No & 1& 2& 3& 4& 5& 6& 7& 8& 9& 10& 11& 12& 13& 14& 15& 16& 17&
18& 19 \\ \hline
$\alpha$ &215$^\circ$& 210$^\circ$& 205$^\circ$& 200$^\circ$& 195$^\circ$& 190$^\circ$& 185$^\circ$& 180$^\circ$& 175$^\circ$& 170$^\circ$&
165$^\circ$& 160$^\circ$& 155$^\circ$& 150$^\circ$& 145$^\circ$& 140$^\circ$& 135$^\circ$& 130$^\circ$& 125$^\circ$ \\ \hline
$\delta_A$ & 4$^\circ$& 4$^\circ$& 4$^\circ$& 4$^\circ$& 7$^\circ$& 9.5$^\circ$& 11.2$^\circ$& 13$^\circ$& 13.75$^\circ$& 15$^\circ$& 16&-&-& 18.4$^\circ$& 19$^\circ$& 19.4$^\circ$& 19.5$^\circ$& 19.5$^\circ$& 19.5$^\circ$ \\ \hline
$\delta_B$ &-&-& 9$^\circ$& 13$^\circ$& 16$^\circ$& 18.5$^\circ$& 20.2$^\circ$& 22$^\circ$& 22.75$^\circ$& 24$^\circ$& 25.5&
26.3$^\circ$& 27.7$^\circ$& 28.4$^\circ$& 29$^\circ$& - & - & - & - \\ \hline
\end{tabular}
\end{center}
\end{table*}

\begin{figure*}
%\begin{center}
\includegraphics[height=4.5cm]{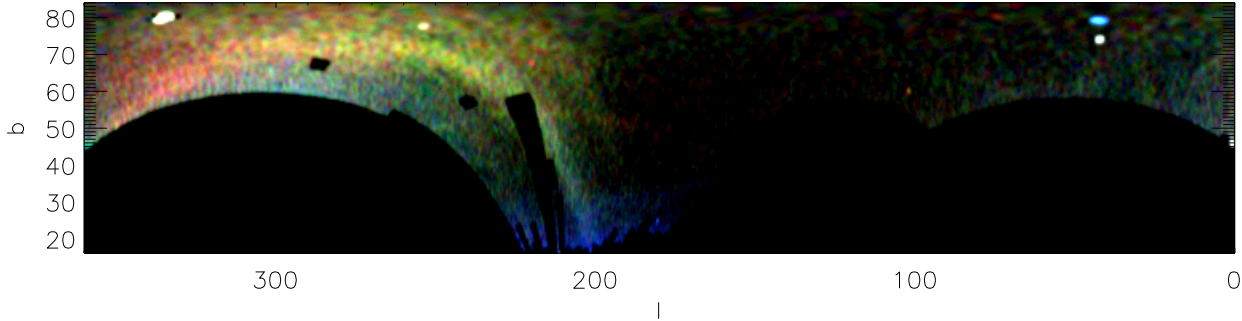}
%\plotone{f3.eps}
\caption{An RGB composite of the SDSS stars as in Figure 1, but now in
Galactic coordinates ($\ell,b$). Note in addition to the branches of
the Sgr stream, a second orphan stream is clearly visible at latitudes
of $b \approx 50^\circ$ and with longitudes satisfying $180^\circ \lta
\ell \lta 230^\circ$. The Monoceros Ring is also discernible at low
latitudes.}
\label{fig:saglb}
%\end{center}
%\vskip-0.5cm
\end{figure*}

\section{The Data and a Simple Color Cut}

SDSS \citep{Yo00} is an imaging and spectroscopic survey that has
mapped $\sim 1/4$ of the sky. Imaging data are produced simultaneously
in five photometric bands, namely $u$, $g$, $r$, $i$, and
$z$~\citep{Fu96,Gu98,Ho01,Am06,Gu06}. The data are processed through
pipelines to measure photometric and astrometric properties
\citep{Lu99,St02,Sm02,Pi03,Iv04} and to select targets for
spectroscopic follow-up.  For de-reddening, we use the maps of
\citet*{Sc98}.  DR5 covers $\sim 8000$ square degrees around the
Galactic North Pole, together with 3 strips in the Galactic southern
hemisphere. We use the catalogue of objects classified as stars with
artifacts removed~\footnote{See
http://cas.sdss.org/astro/en/help/docs/realquery.asp\#flags}, together
with magnitude limits $ r < 22$ and $g < 23$.  At low right ascension
and declination, we are limited by the boundary of DR5. We choose to
cut our sample at $\alpha =230^\circ$ and $\delta=60^\circ$.  This
gives us a total of $\approx 2 \times 10^7$ stars.

Figure~\ref{fig:sag} shows the density of stars satisfying the color
cut $g-r < 0.4$ in SDSS DR5. As we show shortly, these are upper main
sequence and turn-off stars belonging to the Sgr stream.  This is a
RGB-composite image assembled from three grayscale images colored red,
green and blue corresponding to the density of the selected stars in
the following three magnitude bins: red is $21.33 < r \le 22.0$, green
is $20.66 < r \le 21.33$, and blue is $20.0 < r \le 20.66$. For a
given stellar population, these magnitude bins are distance bins, with
red being the most distant and blue the nearest. There is a clear
distance gradient along the stream, from the nearer parts at right
ascensions $\alpha \approx 120^\circ$ to the more distant parts at
$\alpha \approx 210^\circ$. There is also a bifurcation visible in the
stream starting at about the North Galactic Pole; in what follows, we
will refer to the lower declination branch as branch A, and the higher
declination as branch B.

Majewski et al. (2003) traced the Northern stream of the Sgr for right
ascensions $\alpha$ between $270^\circ$ and $190^\circ$. For $\alpha <
190^\circ$, Majewski et al. (2003) did not see a clear continuation of
the stream. The combination of Majewski et al.'s (2003) M giants
together with the SDSS stars in Figure~\ref{fig:sagnew} shows for the
first time the entirety of the stream, including its continuation
through the Galactic Cap and into the Galactic
Plane. Figure~\ref{fig:sagnew} also shows the locations of a number of
globular clusters, some of which are known to be associated with the
Sgr stream. For example, Bellazzini et al. (2003) used 2MASS data to
conclude that NGC 4147 was physically immersed in the stream.

Figure~\ref{fig:sag} displays such a remarkable wealth of Galactic
substructure that it might appropriately be called the ``Field of
Streams''. Among the most visible of these is the whitish-blue
colored, and hence relatively nearby, stellar overdensity centered at
($\alpha \approx 185^\circ, \delta \approx 0^\circ$), found by Juri\'c
et al. (2005) and named the Virgo Overdensity; this is perhaps the
same structure as the nearby 2MASS Northern Fluff (Majewski et
al. 2003). Parts of the Monoceros Ring (Newberg et al. 2002) are
visible as the blue-colored structure at $\alpha \approx 120^\circ$.
Figure~\ref{fig:saglb}, an RGB-composite image of the SDSS stars in
Galactic coordinates ($\ell,b$), also shows the arc-like structures of
the Monoceros Ring, as predicted by the simulations of Pe\~{n}arrubia
et al. (2005).  Two of the globular clusters with tidal tails
previously identified in SDSS data -- namely Pal 5 (Odenkirchen et
al. 2001) and NGC 5466 (Belokurov et al. 2006) -- can be discerned in
the figures, together with their streams. Finally, a new stream is
shown clearly, running from $\alpha,\delta \approx 160^\circ,0^\circ$
to $\alpha,\delta \approx 140^\circ,50^\circ$ ($b \approx 50^\circ$
and $180^\circ \lta \ell \lta 230^\circ$ in
Figure~\ref{fig:saglb}). It is distinct from the Sgr stream, which it
crosses; we discuss its progenitor in a future contribution.
 
%Also,  This map has $720 \times
%140$ pixels, which are $0.5^\circ$ on a side. Also showing clearly in
%this map is a new stream at latitudes of $b \approx 50^\circ$ and with
%longitudes satisfying $180^\circ \lta \ell \lta 230^\circ$. It is
%distinct from the Sgr stream, which it crosses. Its progenitor
%is discussed elsewhere.

%
\begin{figure*}[t]
%\begin{center}
\includegraphics[height=4.8cm]{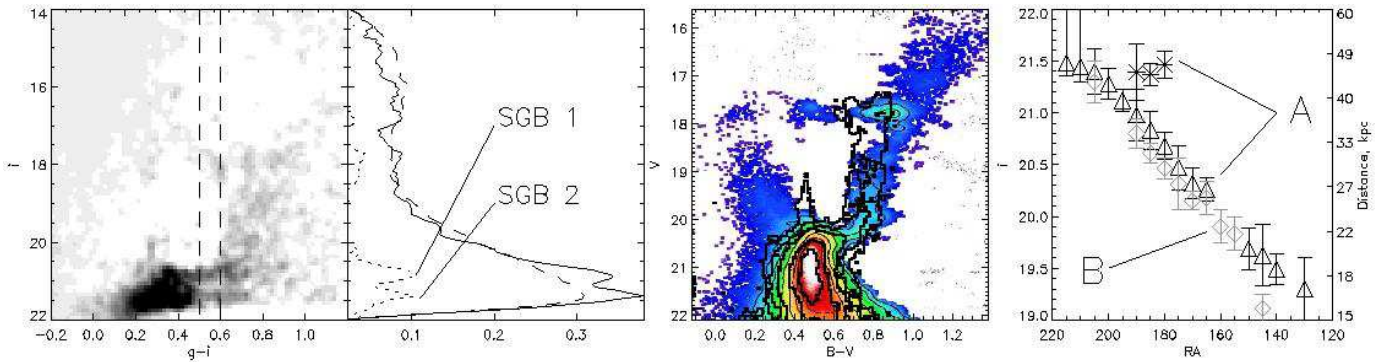}
%\plotone{f4jpg.ps}
\caption{(a): Color-magnitude histogram (Hess diagram) of the
on-stream field A7 (that is, the 7th field on the A branch) minus the
number-scaled corresponding off-stream field. The color bin used to
construct the one-dimensional slice is marked. (b): The luminosity
functions of the one-dimensional slices for the on-stream field
(unbroken), the off-stream field (broken) and the difference
(dotted). Notice the two subgiant peaks corresponding to two
structures at different distances or two distinct stellar
populations. (c): A composite CMD for branch A referenced to field A16
(black contours) overplotted on the CMD from \citet{Be06}
(colored contours) of the body of the Sgr dSph.  (d): Magnitude versus
right ascension of the subgiant branch for the A fields (stars and
triangles) and for the B fields (squares).  The distance in kpc is
given on the right-hand side axis. If a datapoint is missing, a
detection was not possible in that field. }
\label{fig:cmdex}
%\end{center}
%\vskip-0.5cm
\end{figure*}

\section{Tomography of the Sgr Stream}

To analyze the three-dimensional structure of the stream, we set up a
series of $6^\circ \times 6^\circ$ fields along branches A and B,
shown as red (for A) and black (for B) squares in
Figure~\ref{fig:sagnew}. The first three A fields actually probe both
the A and B branches, which are merging at these locations. The
coordinates of the field centers are listed in Table~1. For each
on-stream field, there is a companion off-stream field of size
$15^\circ \times 15^\circ$, which has the same right ascension but is
offset in declination as noted in Table~1. The off-stream fields are
larger, so that the background is as smooth as possible.

Color-magnitude diagrams (CMDs, $g-i$ versus $i$) are constructed for
each of the fields and then normalised by the number of stars. The
difference between each on-stream field and its companion off-stream
field reveals the population of stars belonging to the Sgr stream. An
example for the field A7 is shown in Figure~\ref{fig:cmdex}(a),
together with a one-dimensional slice at $g-i = 0.55$ in
Figure~\ref{fig:cmdex}(b). The subgiant branch is clearly visible and
its location can be found to good accuracy. In fact, we fit a Gaussian
to the one-dimensional slice to obtain its $i$ band magnitude and
uncertainty.  For the example in Figure~\ref{fig:cmdex}(a), there are
two subgiant branches visible in the CMD, and two distinct peaks in
the luminosity functions in Figure~\ref{fig:cmdex}(b). These
correspond to two distinct structures at different distances, possibly
different wraps of the Sgr stream. Although they could correspond to
different populations at the same distance, this seems unlikely as the
magnitude difference between the two subgiant branches changes on
moving along the stream.

The Sgr dSph is known to contain a variety of stellar populations with
different ages and metallicities. The most recent study of this is by
\citet{Be06}, who presented a comprehensive CMD of the body of Sgr.
Rather than attempting to fit multiple theoretical isochrones to our
CMDs, we show a direct comparison between \citet{Be06}'s CMD and a
composite CMD for the whole of the stream. The composite CMD is
created by using the subgiant branch location to measure the magnitude
offset. We use fields A4 to A16, with A16 as the reference. The
result, transformed to $B-V$ versus $V$ \citep{Sm02}, is shown as the
black contours in Figure~\ref{fig:cmdex}(c). This is overlaid upon the
colored contours of Bellazzini et al.'s (2006) CMD, corrected for
extinction in $B-V$ and $V$ by -0.12 and -0.4 magnitudes,
respectively.  The upper main sequence, turn-off, and subgiant
branches are all well-matched.  The only significant difference occurs
in the sparsely populated upper red giant branch (the dark blue
contours). Hence, this stream is entirely consistent with being
composed of the same mix of populations as the Sgr dSph. The 0.6
magnitude offset between our composite CMD and Bellazzini et al.'s is
a direct measure of distance.

Figure~\ref{fig:cmdex}(d) shows the $i$ band magnitude of the subgiant
branch versus right ascension of the fields. The triangles and stars
correspond to the two structures in the A branch, the diamonds to the
B branch. Only one sequence is detected in the B fields. Although its
distance offset is significant only at about the 1 $\sigma$ level, the
B branch is systematically brighter and hence probably slightly
closer.  Assuming the distance to the Sgr dSph is $\sim 25$ kpc, then
Figure~\ref{fig:cmdex}(d) can be converted directly into heliocentric
distance versus right ascension, as shown on the right-hand axis.  The
two distinct structures in the A fields are separated by distances up
to $\approx 15$ kpc. Note that there is no evidence for any part of
the Sgr stream passing close to the Solar neighbourhood, as has
sometimes been conjectured~\citep{Fr04}.

This structure of three branches -- two close together and one more
distant -- is understandable on comparison with the numerical
simulations. For example, the effect is clearly visible in the lower
panel of Figure 3 of Helmi (2004), where there are two close branches
representing material stripped between 3 Gyr and 6.5 Gyr ago, and less
than 3 Gyr ago.  Branches A and B are therefore probably tidal debris,
torn off at different times. The older material, stripped off more
than 6.5 Gyr ago, lies well behind the two close branches in Helmi's
simulations.  This earlier wrap of the orbit may correspond to the
distant structure seen behind the A branch.

\section{Conclusions}

We used a simple color cut $g-r < 0.4$ to map out the distribution of
stars in SDSS DR5. The ``Field of Streams'' is an RGB-composite image
composed of magnitude slices of the stellar density of these stars. It
reveals a super-abundance of Galactic substructure, including the
leading arm of the Sgr stream, as well as a number of sibling streams
(some hitherto unknown).

Part of the Sgr stream has previously been seen in the Northern
hemisphere by Majewski et al. (2003). Here, we have mapped out the
continuation of the stream, as it passes by the Galactic Cap and
returns to the Galactic Plane.  At least two branches of the stream --
labelled A and B -- have been identified, corresponding to material
torn off at different epochs. There is also evidence for a still more
distant structure behind the A branch.

The Sgr stream provides a probe of the shape of the Galactic
halo. Previous work has been hampered by the absence of data in the
most important regions of the sky. For example, \cite{He04} provides
simulations of the Sgr stream for a range of halo shapes from extreme
oblate to prolate, all of which broadly agree with the data available
at that time. Our data covers the critical region where the models
differ substantially.  Examination of the simulations displayed in
Figure 2 of \citet{He04} enables some preliminary conclusions to be
drawn. In oblate haloes, the Sgr stream is much fatter and the stars
much more scattered than shown in our Figure 1. Only the spherical and
very mildly prolate haloes ($1 \lta q < 1.1$, where $q$ is the axis
ratio in the potential) seem reasonable matches to the data. In both
these simulations, the Sgr stream shows the same bifurcation and
overall morphology as in the SDSS data. We suggest that this may be a
powerful discriminant of halo shape (Fellhauer et al. 2006).

\acknowledgments 
We thank M. Bellazzini and the anonymous referee for helpful discussions.
Funding for the SDSS and SDSS-II has been provided by the Alfred P.
Sloan Foundation, the Participating Institutions, the National Science
Foundation, the U.S. Department of Energy, the National Aeronautics
and Space Administration, the Japanese Monbukagakusho, the Max Planck
Society, and the Higher Education Funding Council for England. The
SDSS Web Site is http://www.sdss.org/.
                                                                               
The SDSS is managed by the Astrophysical Research Consortium for the
Participating Institutions. The Participating Institutions are the
American Museum of Natural History, Astrophysical Institute Potsdam,
University of Basel, Cambridge University, Case Western Reserve
University, University of Chicago, Drexel University, Fermilab, the
Institute for Advanced Study, the Japan Participation Group, Johns
Hopkins University, the Joint Institute for Nuclear Astrophysics, the
Kavli Institute for Particle Astrophysics and Cosmology, the Korean
Scientist Group, the Chinese Academy of Sciences (LAMOST), Los Alamos
National Laboratory, the Max-Planck-Institute for Astronomy (MPIA), the
Max-Planck-Institute for Astrophysics (MPA), New Mexico State
University, Ohio State University, University of Pittsburgh,
University of Portsmouth, Princeton University, the United States
Naval Observatory, and the University of Washington.

\end{document}